\documentclass[cits]{PoS}

\usepackage{amsmath}

\title{Magnetic properties of the neutron in a uniform background field}

\ShortTitle{Magnetic properties of the neutron in a uniform background field}

\author{\speaker{Thomas Primer}\\%
        Centre for the Subatomic Structure of Matter (CSSM), School of Chemistry and Physics, University of Adelaide, SA 5005, Australia\\
        E-mail: \email{thomas.primer@adelaide.edu.au}}
      	
\author{{Waseem Kamleh}\\
		Centre for the Subatomic Structure of Matter (CSSM), School of Chemistry and Physics, University of Adelaide, SA 5005, Australia\\
        E-mail: \email{waseem.kamleh@adelaide.edu.au}}
		
\author{Derek Leinweber\\
		Centre for the Subatomic Structure of Matter (CSSM), School of Chemistry and Physics, University of Adelaide, SA 5005, Australia\\
        E-mail: \email{derek.leinweber@adelaide.edu.au}}
		
\author{Matthias Burkardt\\
        Department of Physics, New Mexico State University, Las Cruces, NM 88003-8001, USA\\
        E-mail: \email{burkardt@nmsu.edu}}

\abstract{We present calculations of the magnetic moment and magnetic polarisability of the neutron from the background field method.
  The calculations are performed on $32^3\times64$ dynamical lattices generated by the PACS-CS collaboration and made available via the ILDG.
  We consider uniform fields quantised by the periodic spatial volume.
  We explore different approaches for improving the quality of the fits used in the results.
  Also included are initial results for the magnetic moment of the lowest lying negative parity nucleon states.}

\FullConference{The 30th International Symposium on Lattice Field Theory\\
		 June 24 - 29,  2012\\
		 Cairns, Australia}

\begin{document}

\section{Introduction}
	The magnetic moment and magnetic polarisability are fundamental properties of a particle which describe its response to the application of an external magnetic field.
These properties can be calculated from first principles using lattice QCD techniques.
One method is to use three-point functions with an explicit current in order to calculate electromagnetic form factors, from which the moment can be derived \cite{Leinweber:1991,Boinepalli:2006}.
Here we present results from the background field method \cite{Draper:1982,Burkardt:1996,LeePol:2005,Lee:2008,Alexandru:2008}, which involves adding a phase factor to the gauge field which induces a constant magnetic field across the whole lattice.
This field causes a shift in the energy of the particle which can be measured and used to calculate both the magnetic moment and magnetic polarisability via the energy-field relation
	\begin{equation}\label{erelation}
	E(B) = M_N + \vec\mu \cdot \vec{B} + \frac{e\vert{B}\vert}{2M_N} - \frac{4\pi}{2}\beta B^2 + \mathcal{O}(B^3).
	\end{equation}
	
	The phase factor used to apply the magnetic field has the exponential form $\exp{(iaqA_\mu(x))}$.
This leads to a quantisation condition on the available choices for the magnitude of the magnetic field.
If the size of the lattice is too small then the field will be large and the energy-field relation will no longer be dominated by the leading terms.
Previous calculations using this technique have avoided the problem of the quantisation condition by using a linearised version of the phase factor with Dirichlet boundary conditions \cite{LeePol:2005,Lee:2008}, however this introduces new finite volume errors.
The results presented here are calculated with the full exponential phase factor on a periodic lattice.
These results include the magnetic moment and magnetic polarisability of the neutron, as well as a preliminary look at the magnetic moment of the lowest lying negative parity nucleon states.
	
\section{The Background Field Method}
	To add a background magnetic field to lattice calculations we begin by considering the continuum case, modifying the covariant derivative with the addition of a minimal electromagnetic coupling,
\begin{equation}
D_{\mu} = \partial_{\mu}+gG_{\mu}+qA_{\mu},
\end{equation}
where $A_{\mu}$ is the electromagnetic four-potential and $q$ is the charge on the fermion field.
On the lattice this is equivalent to multiplying the usual gauge links by a simple phase factor
\begin{equation}
U_{\mu}^{(B)}(x)=\exp(iaqA_{\mu}(x)).
\end{equation}

To obtain a uniform magnetic field along the z-axis we note that $\vec{B} = \vec{\nabla} \times \vec{A},$ and hence 
\begin{equation}\label{Bz}
B_z = \partial_x A_y - \partial_y A_x.
\end{equation}
Note that this equation does not specify the gauge potential
uniquely. There are multiple valid choices of $A_\mu$ that give rise
to the same field, however they are equivalent up to a gauge
transformation. We choose $A_x(x,y) = By$ to produce a constant
magnetic field of magnitude $B$ in the z direction.  Examining a
single plaquette in the $(\mu,\nu) = (x,y)$ plane shows that this
gives the desired field for a general point on the lattice.
However on a finite lattice $(0 \le x \le N_x-1), (0 \le y \le N_y-1)$
there is a discontinuity at the boundary due to the periodic boundary
conditions.  In order to fix this problem we make use of the
$\partial_x A_y$ term from Eq.~\eqref{Bz}, giving $A_y$ the
following values,
\begin{equation}
A_y(x,y) = \begin{cases} 0 & \mathrm{for\ } y < N_y-1 \\
	           -N_y Bx & \mathrm{for\ } y = N_y-1, \end{cases}
\end{equation}
such that we now get the required value at the $y=N_y-1$ boundary.

This still leaves the double boundary, $x = N_x-1$ and $y=N_y-1$, where the plaquette only has the required value under the condition
$\exp(-ia^2 qBN_x N_y)=1$.
This gives rise to the quantisation condition which limits the choices of magnetic field strength based on the lattice size
\begin{equation} \label{quantisationcondition}
a^2 qB = \frac{2\pi n}{N_x N_y},
\end{equation}
where $n$ is an integer specifying the field strength in multiples of the minimum field strength quantum.

\section{Simulation Details}

The results presented here were calculated on 2+1 flavour dynamical-QCD gauge field configurations on $32^3 \times 40$ lattices. These were provided by the PACS-CS collaboration \cite{PACS-CS} as part of the International Lattice Data Grid (ILDG) \cite{ILDG}. They have a physical lattice spacing of $a=0.0907$ fm and a beta value of $\beta = 1.9$. There are five values of the light quark mass corresponding to the pion masses $m_\pi =$ 622, 512, 388, 282, 151 MeV, however the lightest mass has not yet been considered herein. It should be noted that the configurations are dynamical only in the QCD sense, they have not been calculated with the background field included as this is prohibitively expensive. The corrections due to this are expected to be small.
Unless mentioned otherwise correlation functions were calculated using the interpolating field $\chi_1=(u^TC\gamma_5d)u$ with 100 sweeps of stout smearing at the source.

\section{Magnetic Moment}

Once the background field method has been used to calculate correlation functions with a magnetic field on the lattice we use the energy relation of Eq.~\eqref{erelation} to determine the magnetic moment.
The term we want to isolate is $\vec\mu \cdot \vec{B}$, which we can do by taking advantage of the fact that the sign of this term is changed when the spin of the particle is flipped, while the magnitude stays the same. We simply take the difference of spin-up and spin-down energies,
\begin{equation}
\frac{1}{2}(E_\uparrow - E_\downarrow) = \vec\mu \cdot \vec{B},
\end{equation}
which is done by constructing the following ratio of correlation functions,
\begin{equation}
\delta E(B) = \frac{1}{2}\left(\ln\left(\frac{G_{\uparrow}(B,t)}{G_{\uparrow}(0,t)}\frac{G_{\downarrow}(0,t)}{G_{\downarrow}(B,t)}\right)\right)_{\rm fit},
\end{equation}
leaving us with only the term we are interested in. Although including the zero field correlators is not strictly necessary, it is useful for allowing correlated errors to cancel as much as possible.

\begin{figure}
\centering
\includegraphics[width=0.35\textwidth,angle=90]{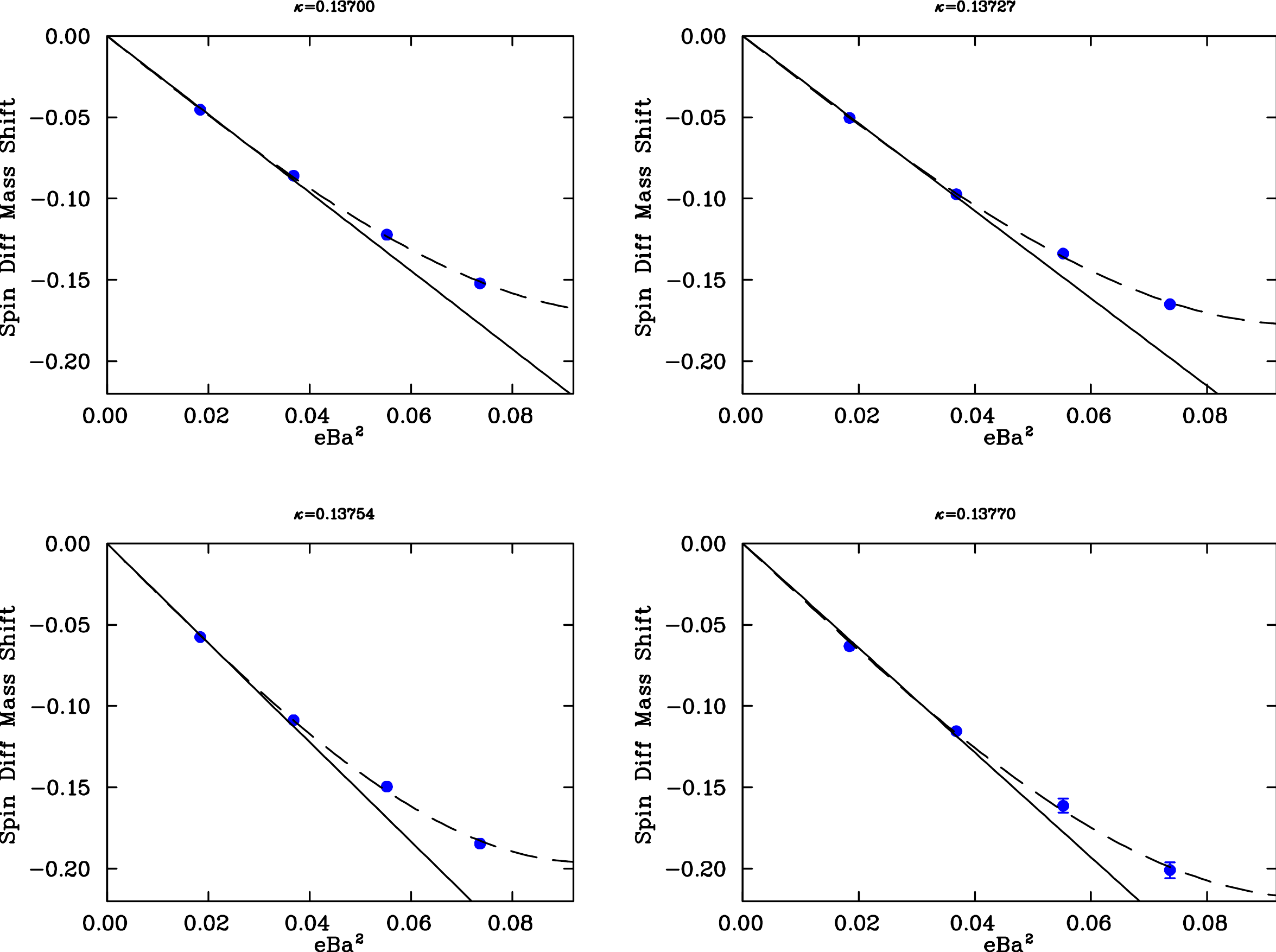}\hspace{12pt}
\includegraphics[width=0.35\textwidth,angle=90]{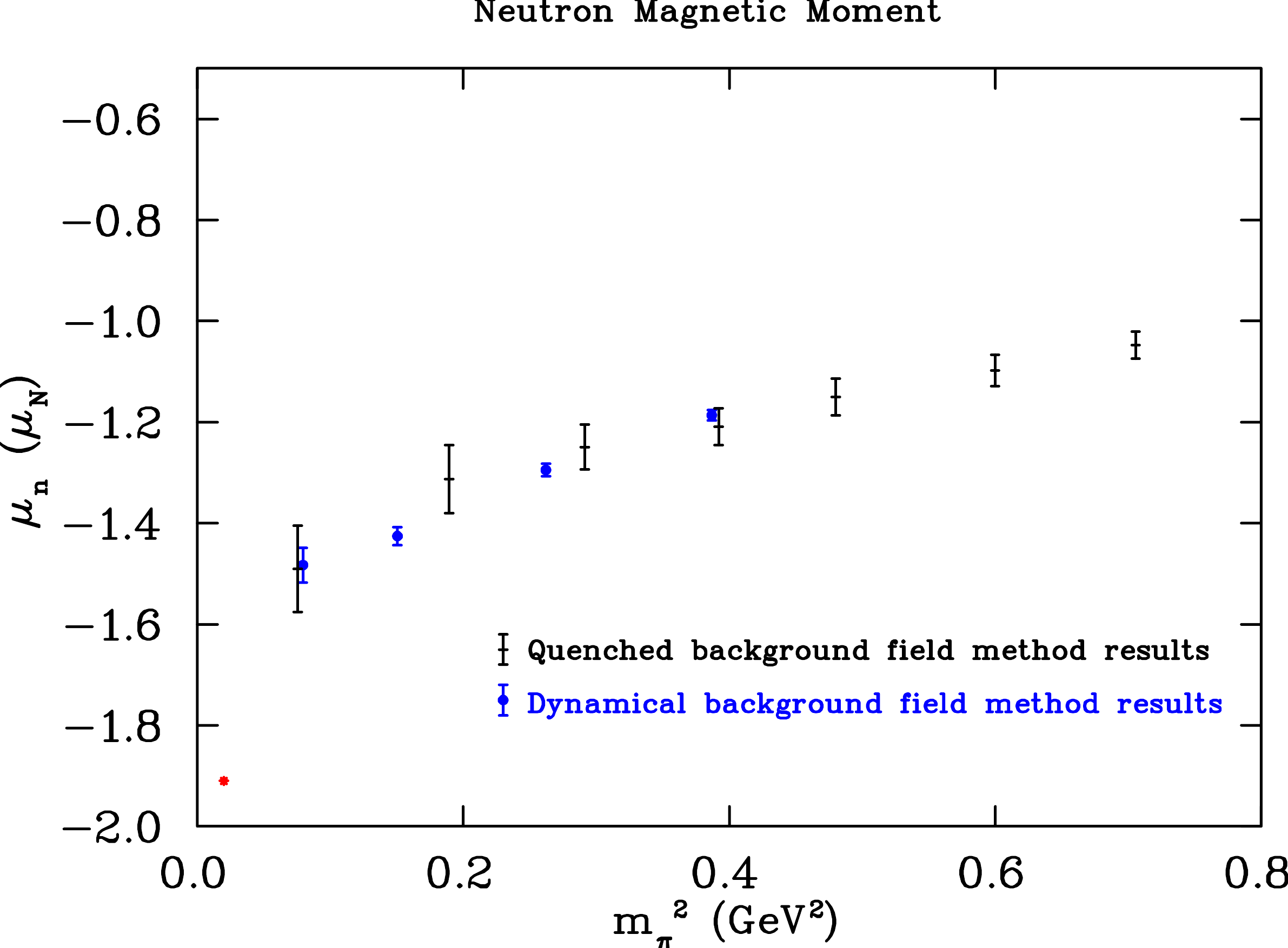}
\caption{Left: Plot of the field strength vs the spin-difference mass shift. The solid line is a purely linear fit to the first two point and the dashed line is a linear plus cubic fit to all four points. Right: Plot of the magnetic moment vs pion mass squared showing a comparison of these results with older, quenched results. The left-most point marks the experimental value \cite{PDG:2010}.}
\label{diffplots}
\end{figure}

Figure \ref{diffplots} shows fits of the spin-difference mass shift to the field strength.
These fits give the value of the magnetic moment from the linear coefficient.
Both a purely linear fit to only the first two field strengths and a linear plus cubic term fit to all field strengths were calculated.
This was to ensure that the fields used are small enough for the energy field relation \eqref{erelation} to be valid.
Since the linear terms from each fit agree well within errors we can be confident that higher order terms are not unduly affecting the result.

Figure \ref{diffplots} also gives the plot of the neutron magnetic moment results.
Comparing quenched and dynamical values we see that they agree within errors and are both trending generally towards the physical value.
Once the lightest mass has been calculated the trend should become clear and it will be relatively easy to do a chiral extrapolation to the physical pion mass and infinite volume.

\subsection{Negative Parity Nucleon Magnetic Moment}
It is also possible to examine the lowest lying negative parity nucleon state using the same correlation functions as for positive parity.
This can be seen by writing the correlation function as a sum over positive and negative parity baryon states,
\begin{equation}
  G(t,\vec{p}) = \sum_{B^+}\lambda_{B^+} \bar{\lambda}_{B^+}e^{-E_{B^+}t}\frac{\gamma\cdot p_{B^+}+M_{B^+}}{2E_{B^+}} - \sum_{B^-}\lambda_{B^-} \bar{\lambda}_{B^-}e^{-E_{B^-}t}\frac{-\gamma\cdot p_{B^-}+M_{B^-}}{2E_{B^-}}.
\end{equation}
Then after setting $\vec{p} = 0$ to give $E_{B^\pm}=M_{B^\pm}$ the negative parity spin-up and spin-down parts can be projected into the (3,3) and (4,4) components of the Dirac matrix via the parity projector, $\Gamma_\pm = (1\pm\gamma_0)/2$.
\begin{figure}
\centering
\includegraphics[width=0.35\textwidth,angle=90]{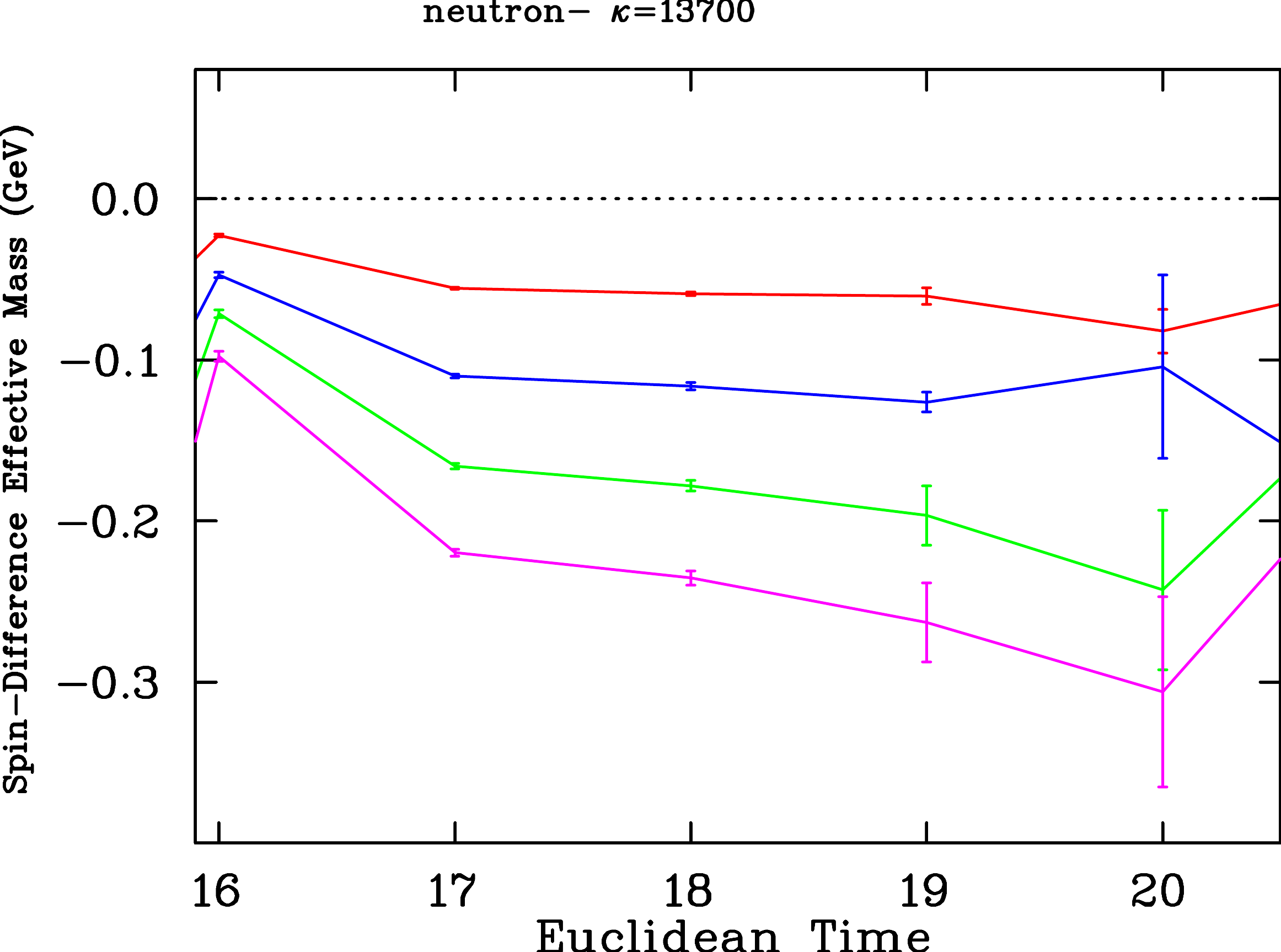}\hspace{12pt}
\includegraphics[width=0.35\textwidth,angle=90]{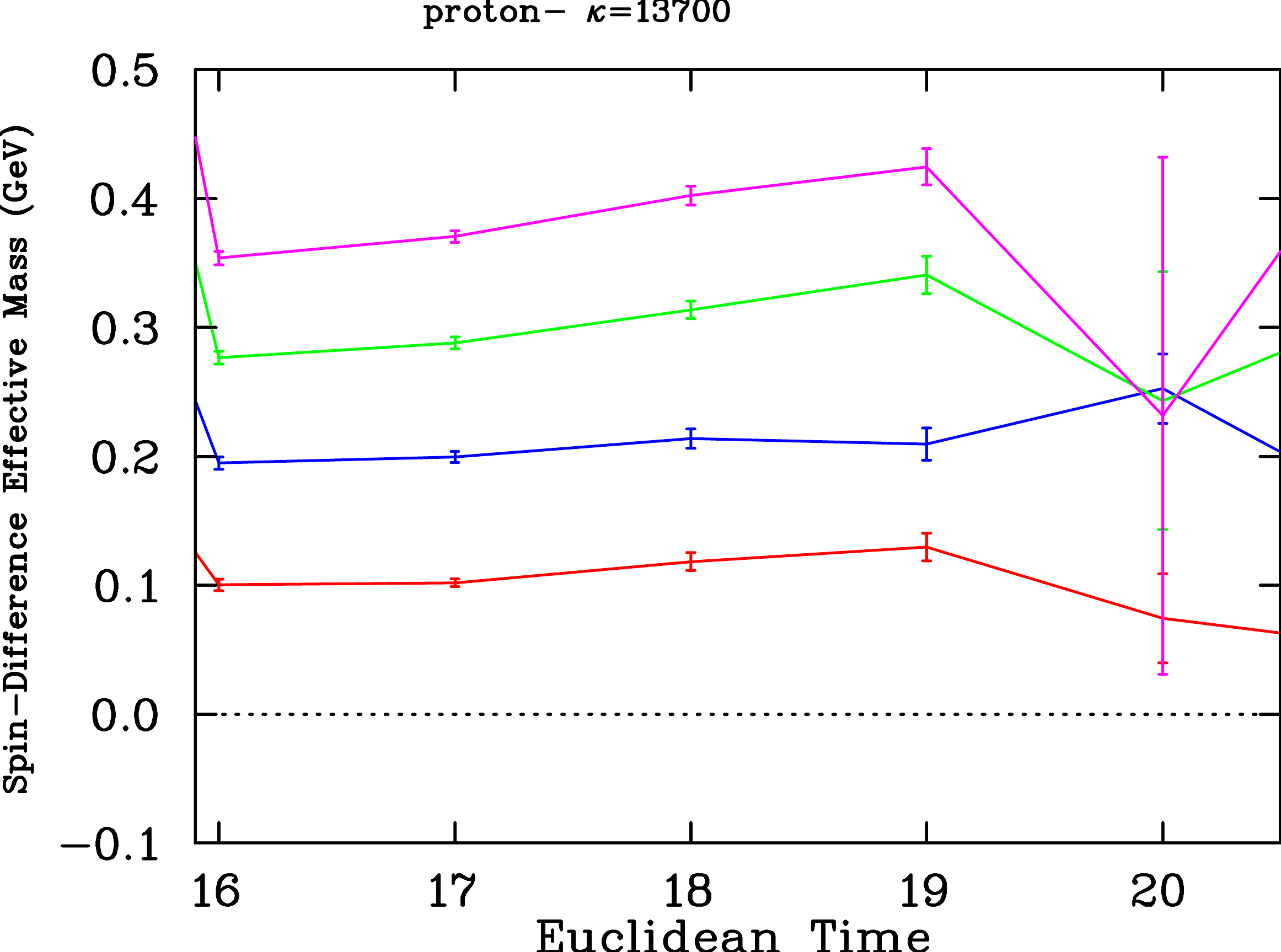}
\caption{Spin-difference effective mass plots for the negative parity proton and neutron using $\chi_2$ interpolating fields at the heaviest quark mass with all four fields strengths.}
\label{npplots}
\end{figure}

Effective mass plots for the standard $\chi_1$ type interpolating field showed poor signal.
Therefore we decided to try using the interpolating field type $\chi_2= (q^T\,C\,q)\,\gamma_5\, q$.
Figure \ref{npplots} shows spin-difference effective mass plots for the neutron and the proton using a $\chi_2$ interpolating field.
Each one shows clear signal and an obvious linear progression over the fields as expected from the magnetic moment effect.

Whilst these plots appear to give fairly clear results for the magnetic moment, we must be careful.
The two lowest lying negative parity nucleon states are the the N(1535) and the N(1650).
The difference in mass between them is quite small, therefore we expect a lot of mixing.
This means we cannot designate a specific state to each of the effective mass plots.
In order to differentiate these states a variational analysis of a correlation matrix will be required, after which a real result for the negative parity nucleon magnetic moments can be reported.

\section{Magnetic Polarisability}

Similar to the magnetic moment case, calculating the magnetic polarisability requires isolating the desired term in the energy relation of Eq.~\eqref{erelation}.
This time we take the average of spin-up and spin-down and also subtract the zero field energy in order to remove the bare mass,
\begin{equation}
\frac{1}{2}((E_\uparrow - M_\uparrow) + (E_\downarrow - M_\downarrow)) = \frac{e\vert{B}\vert}{2M_N} - \frac{4\pi}{2}\beta B^2,
\end{equation}
which in terms of correlation functions looks like,
\begin{equation}
\delta E(B) = \frac{1}{2}\left(\ln\left(\frac{G_{\uparrow}(B,t)}{G_{\uparrow}(0,t)}\frac{G_{\downarrow}(B,t)}{G_{\downarrow}(0,t)}\right)\right)_{\rm fit}.
\end{equation}
In addition to the required magnetic polarisability term there is also another term left over. This term is the Landau energy and cannot be isolated from the polarisability term. Fortunately this term is zero for the neutron due to its lack of overall charge, so it can be ignored in this case.

\begin{figure}
\centering
\includegraphics[width=0.35\textwidth,angle=90]{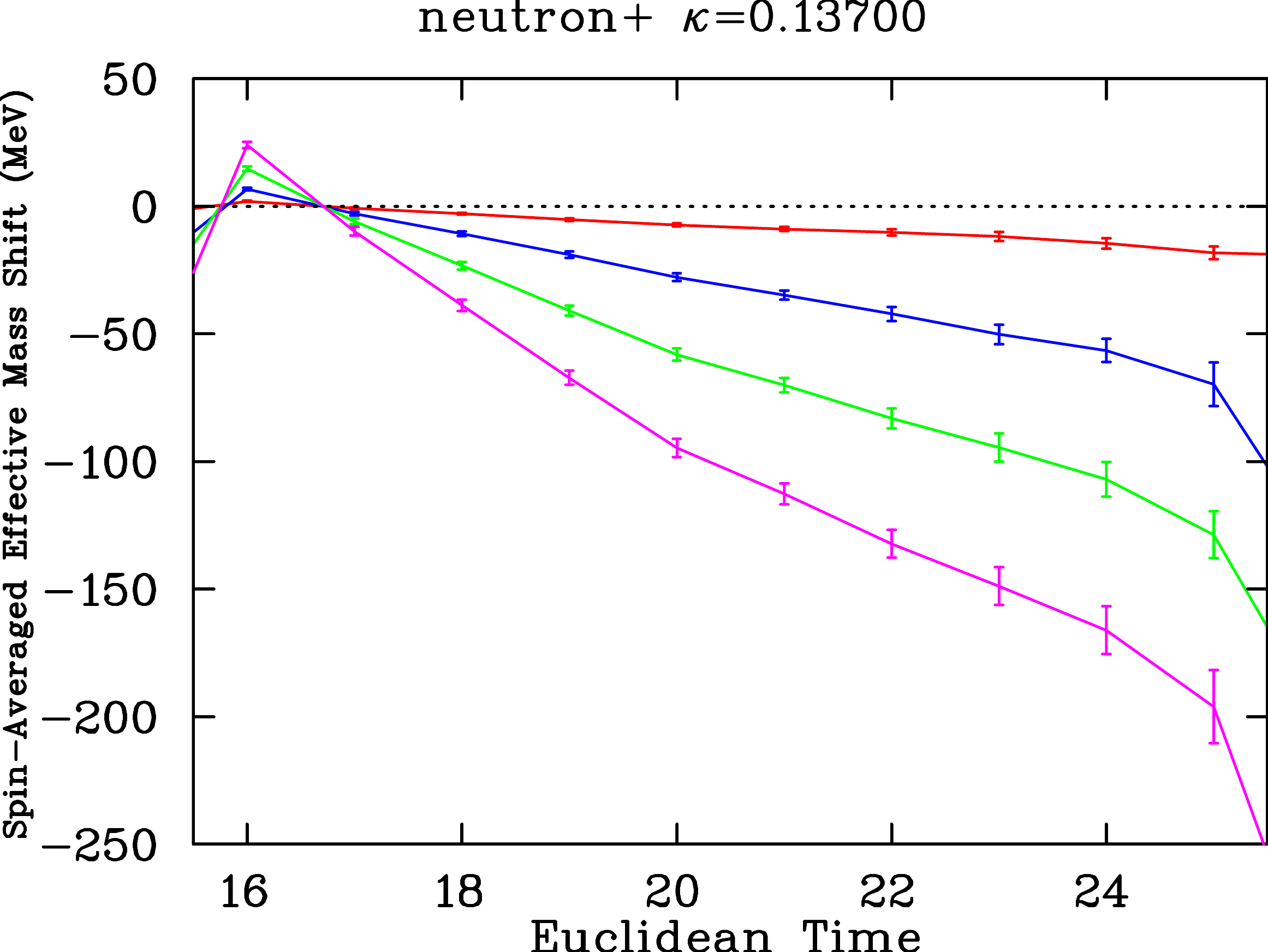}\hspace{12pt}
\includegraphics[width=0.35\textwidth,angle=90]{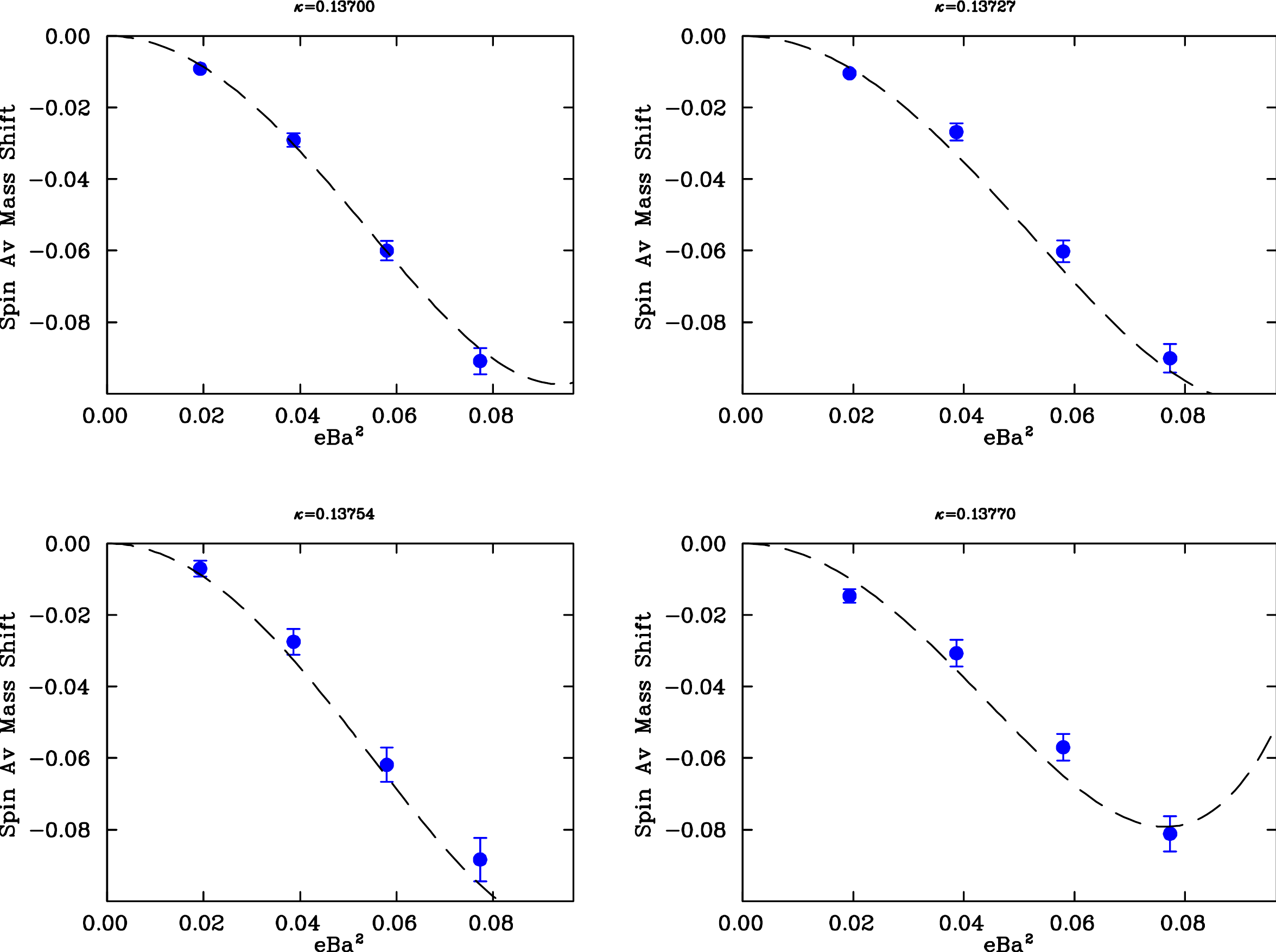}
\caption{Left: The spin-averaged effective mass shifts for the heaviest quark mass at all four field strengths. Right: Quadratic plus quartic fits of the mass shift to the field strength.}
\label{sumplots}
\end{figure}

Even without any Landau energy to worry about the magnetic polarisability is significantly harder to extract than the moment.
First of all the effect is much smaller at the important field strengths.
Secondly the energy shift due to the polarisability arises because the presence of the field deforms the hadron away from its usual spherical shape.
Since it is typical to use spherically smeared sources the creation operators have poor overlap with polarised states and it takes time for the deformation to appear.
As seen in the left hand plot of Fig.~\ref{sumplots} it is possible for the signal to start to become noisy before the energy reaches a proper plateau, making it difficult to fit.

\begin{figure}
\centering
\includegraphics[width=0.35\textwidth,angle=90]{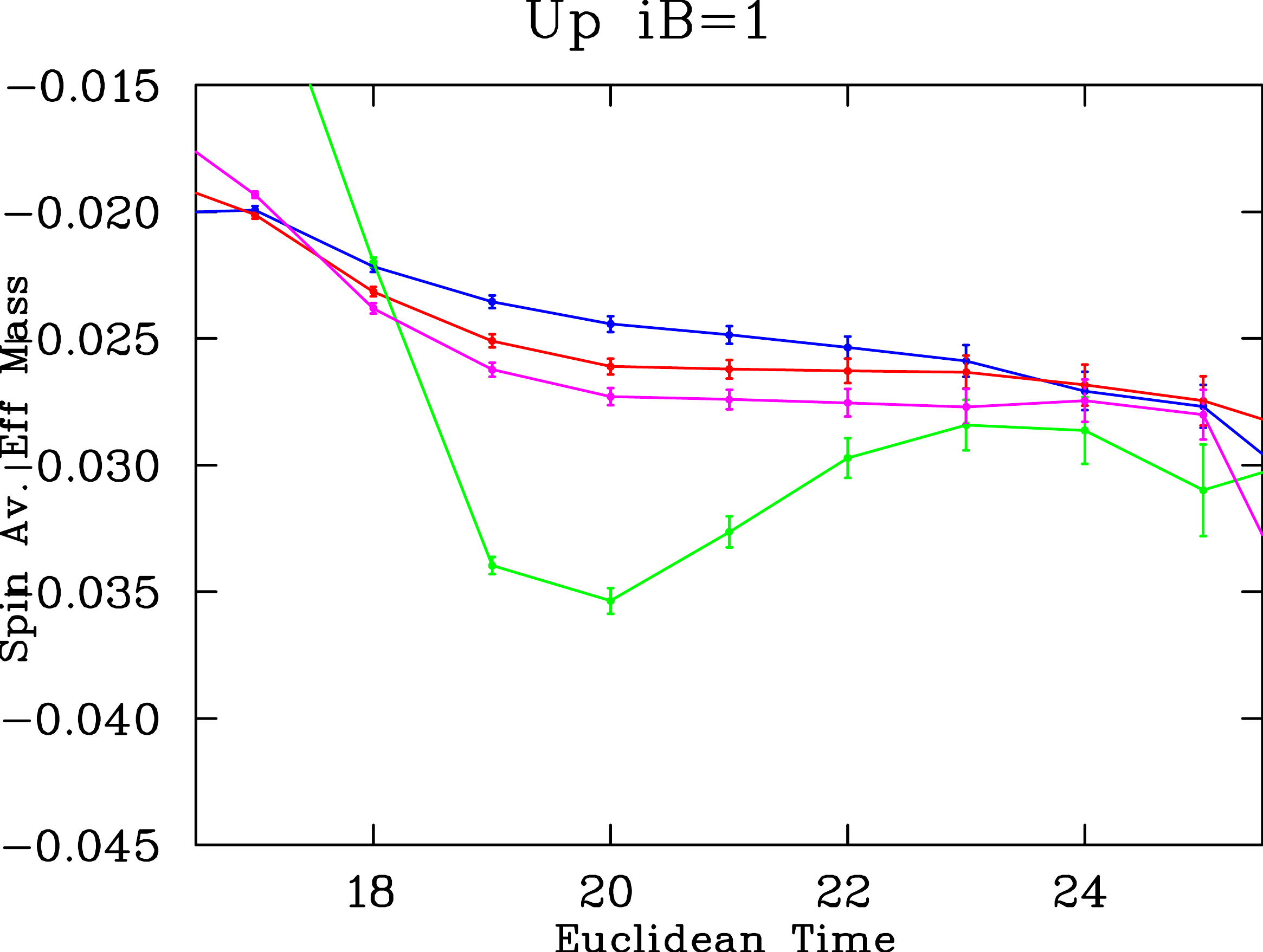}\hspace{12pt}
\includegraphics[width=0.35\textwidth,angle=90]{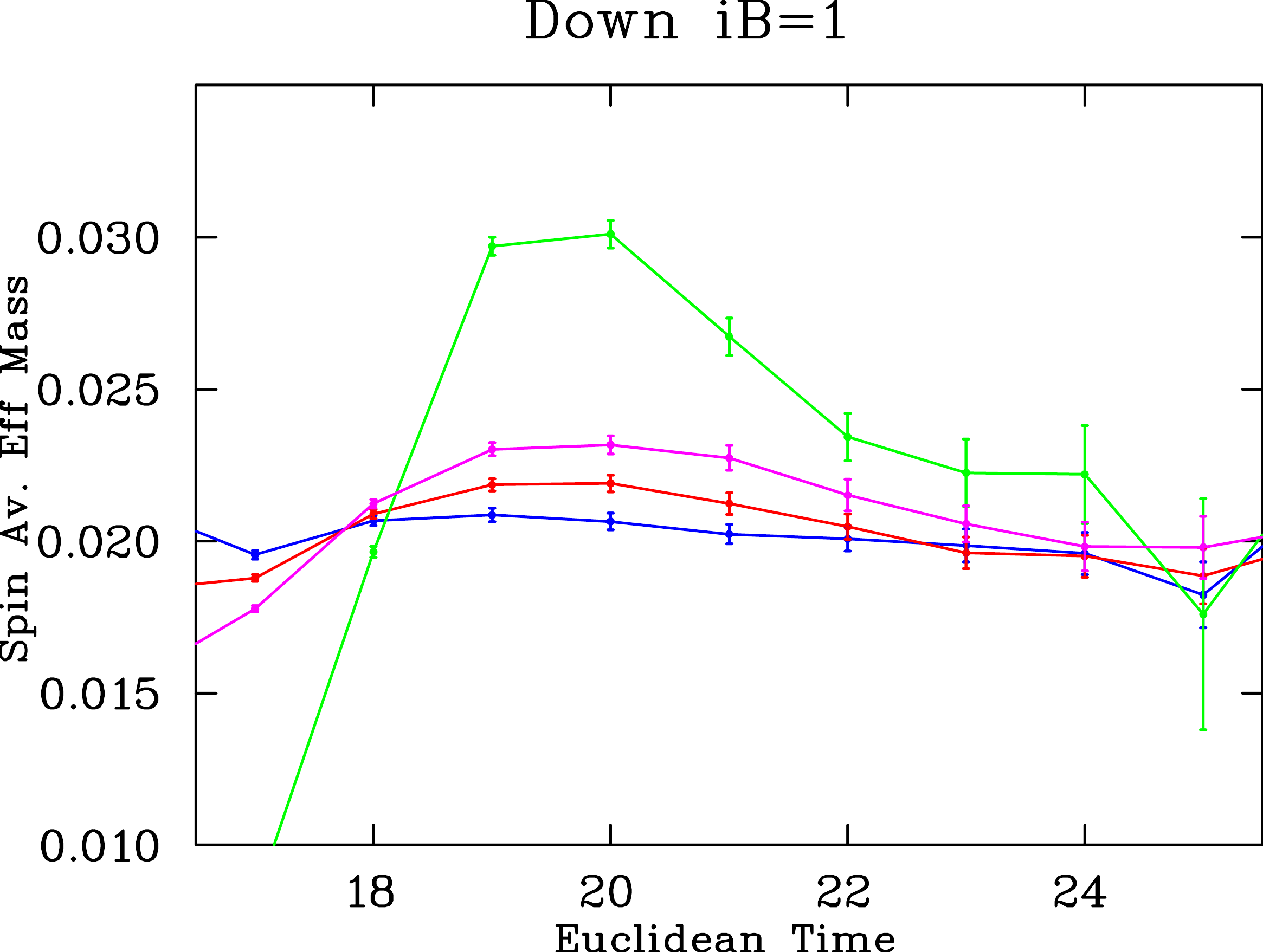}
\caption{The shift in the effective mass due to the field for spin-up and spin-down at the heaviest quark mass and smallest non-zero field strength. For spin-up the lines are, from top to bottom, 100 sweeps of smearing, 35 sweeps, 16 sweeps and zero sweeps. For spin-down the order is reversed.}
\label{smearplots}
\end{figure}

In order to get better plateau behaviour we examined different levels of smearing, looking at 100, 35 and 16 sweeps of smearing.
We also tried a point source in the hope that removing all bias towards any shape would allow it to conform more easily to the final shape of a polarised neutron.
Figure \ref{smearplots} shows plots of the energy shift due to the field for spin-up and spin-down.
For each spin the point source has significant excited state contamination, leading to poor plateaus.
The smeared sources all give similar results at large time, indicating the isolation of the ground state just prior to signal loss.

It can be seen that the best plateau behaviour for spin-up comes from the 16 sweeps of smearing, whilst the best behaviour for spin-down comes from 100 sweeps.
In order to take advantage of this difference we combined the spin-up 16 sweeps of smearing correlation functions with the spin-down 100 sweeps of smearing ones to create a new spin-average.
Whilst the combination had slightly better plateau behaviour the improvement was not very significant, but it may be possible to improve the result further using a variational analysis of multiple smearings in order to pick out the best parts of each one. Another approach would be to use anisotropic smearing to try and match the shape of the polarised wavefunction, however this would likely require significant tuning.

\section{Conclusion}

We have presented a variety of calculations of magnetic properties of the neutron using the background field method with a uniform background field.
Results for the magnetic moment are clear and precise.
There is good agreement between these dynamical results and previous quenched results and a good approach to the physical value.
Magnetic polarisability results are more difficult but still promising.
We have examined ways in which to improve the plateaus needed to fit for the polarisability and hope to improve these further using more advanced techniques.
We have performed preliminary calculations of the magnetic moment of negative parity nucleon states.
We found strong signal using the $\chi_2$ interpolating field but require a more detailed analysis in order separate the two lowest lying negative parity states.
Overall the background field method continues to be a valuable tool for investigating the magnetic properties of hadrons.

\end{document}